\documentclass[aps,pre,twocolumn,amsmath,amsfonts,showpacs]{revtex4}
\usepackage{graphicx}

\begin{document}

\title{Pattern selection in parametrically-driven arrays of nonlinear
  resonators}
\author{Eyal Kenig}
\affiliation{Raymond and Beverly Sackler School of Physics and
  Astronomy, Tel Aviv University, Tel Aviv 69978, Israel}
\author{Ron Lifshitz}
\email[Corresponding author:\ ]{ronlif@tau.ac.il}
\affiliation{Raymond and Beverly Sackler School of Physics and
  Astronomy, Tel Aviv University, Tel Aviv 69978, Israel}
\author{M.~C.~Cross}
\affiliation{Condensed Matter Physics 114-36, California Institute of
  Technology, Pasadena, California 91125}

\date{August 26, 2008}

\begin{abstract}
  We study the problem of pattern selection in an array of
  parametrically-driven nonlinear resonators with application to
  microelectromechanical and nanoelectromechanical systems (MEMS \&
  NEMS), using an amplitude equation recently derived by Bromberg,
  Cross, and Lifshitz [Phys.\ Rev.\ E {\bf 73}, 016214 (2006)]. We
  describe the transitions between standing-wave patterns of different
  wave numbers as the drive amplitude is varied either
  quasistatically, abruptly, or as a linear ramp in time. We find
  novel hysteretic effects, which are confirmed by numerical
  integration of the original equations of motion of the interacting
  nonlinear resonators.

\end{abstract}

\maketitle

\section{Introduction}

Interest in the nonlinear dynamics of microelectromechanical and
nanoelectromechanical systems (MEMS \& NEMS) has grown rapidly over
the last few years, driven by a combination of practical needs as well
as fundamental questions~\cite{LCreview}. Lithographic fabrication
technology allows the construction of large arrays of MEMS \& NEMS
devices (as many as 2800 to date~\cite{Bargatin08}), coupled by
electric, magnetic, or elastic forces. In addition, nonlinear behavior
is readily observed in these devices at relatively small amplitudes
of motion~\cite{turner98,C00,BR01,blick02,turner02,turner03,yu02,cleland04,%
  erbe00,kozinsky07,demartini07,masmanidis07}. Limitations in the
fabrication technology mean that individual devices will usually have
slightly different resonant frequencies, and nonlinear collective
effects, such as synchronization (all devices oscillating in
phase)~\cite{sync1,sync2} and pattern formation~\cite{BR,sato06,LC,BCL}
(coherent response with a more complex spatial structure), have been
proposed as ways of achieving useful coherent responses.
Consequently, for many technological applications, there exists a
practical need to understand the collective nonlinear behavior of MEMS
\& NEMS devices.

At the same time, the advances in the fabrication, transduction, and
detection of MEMS \& NEMS resonators opens up an exciting new
experimental window into the study of fundamental questions in
collective nonlinear dynamics. Typical nonlinear MEMS \& NEMS
resonators are characterized by extremely high frequencies---recently
going beyond 1 GHz~\cite{HZMR03,cleland:070501}---and relatively weak
dissipation, with quality factors in the range of $10^2-10^5$. For
such devices, transients die out rapidly, so that it is easy to attain
the long-time asymptotic states, be they steady, periodic, or chaotic,
and to acquire sufficient data to characterize these states well. From
the theoretical point of view, the systems have the advantage that the
basic physics of the individual elements is simple, and the parameters
can be measured or calculated, so that the equations of motion
describing the system can be established with confidence.  This, and
the fact that weak dissipation can be treated as a small perturbation,
provide a great advantage for quantitative theoretical study.
Moreover, the ability to fabricate arrays of thousands of coupled
resonators opens new possibilities in the study of nonlinear dynamics
of intermediate numbers of degrees of freedom---much larger than one
can study in macroscopic or table-top experiments, yet much smaller
than one studies when considering nonlinear aspects of phonon dynamics
in a crystal.

Our current studies are motivated by the experimental work of Buks and
Roukes~\cite{BR}, who fabricated an array of nonlinear micromechanical
doubly-clamped gold beams, and excited them parametrically by
modulating the strength of an externally-controlled electrostatic
coupling between neighboring beams. The Buks and Roukes experiment was
modeled by Lifshitz and Cross~\cite{LC} using a set of coupled
nonlinear equations of motion. They used secular perturbation theory
to convert these equations of motion into a set of coupled nonlinear
\textit{algebraic\/} equations for the normal mode amplitudes of the
system, enabling them to obtain exact results for small arrays, but
only a qualitative understanding of the dynamics of large arrays. In
order to obtain analytical results for large arrays, Bromberg, Cross,
and Lifshitz~\cite[henceforth BCL]{BCL} studied the same system of
equations, approaching it from the continuous limit of infinitely-many
degrees of freedom, and obtaining a description of the slow
spatiotemporal dynamics of the array of resonators in terms of an
amplitude equation. BCL showed that this amplitude equation could
predict the initial mode that develops at the onset of parametric
oscillations as the driving amplitude is gradually increased from
zero, as well as a sequence of subsequent transitions to other
single-mode oscillations.

The combination of many degrees of freedom and nonlinearity in the
equations of motion typically leads to a large multiplicity of
physically realizable solutions for fixed system parameters. This is
illustrated for the particular case of two and three parametrically
driven oscillators by the explicit results of Lifshitz and
Cross~\cite{LC}. The richness of possible solutions leads to
opportunities for diverse functionality of the system, in nature or
technology. On the other hand we need to be able to predict which out
of the possible solutions will be seen for a given experimental
protocol, or design particular protocols such that the desired
solution is the one that is formed. This is the general question of
\emph{pattern selection}~\cite{reviewcross}. A common experimental
protocol is to vary one or more system control parameters, usually
either slowly compared with the intrinsic time scales of the
dynamics, or in an abrupt step. A particular solution will usually
survive (evolving adiabatically in the former case of slow parameter
variation) until it becomes unstable to small perturbations, and a
sequence of patterns can be predicted by analyzing these
instabilities.

In this paper we investigate the sequence of single mode standing
wave patterns to be expected in parametrically driven oscillator
arrays, in cases where many such modes are simultaneously stable,
when the strength of the driving is varied. Although the quantitative
analysis could be done directly from the basic oscillator equations
of motion, it is advantageous to formulate the analysis in terms of
the BCL amplitude equation. This allows us to display the range of
stable patterns on a reduced plot involving just two dimensionless
variables (a scaled measure of the driving strength, and a scaled
mode wave number), so that it is easy to deduce the general
qualitative behavior on varying parameters. The specific quantitative
behavior for a physical system is also easy to obtain by evaluating
the corresponding scaled quantities. A change of pattern occurs when
parameters vary so that the mode moves outside of the region of
stable patterns on this plot, and the new pattern is predicted by
analyzing the result of the instability using the BCL amplitude
equation. This type of approach has been used in other pattern
forming systems~\cite{ksz}. A novel feature of the present system is
that the difference in the instabilities encountered on increasing
and decreasing the (scaled) driving strength leads to the prediction
of quite different sized mode jumps for the up and down sweeps.

The outline of the paper is as follows. In Sec.~\ref{sec:bcl} we
review the derivation of the amplitude equation of BCL, and in
Sec.~\ref{sec:single} use this equation to discuss the stability of
single-mode oscillating patterns. We then study the sequence of
patterns observed for a variety of time dependent sweeps of the
driving strength: quasistatic variation in Sec.~\ref{quasi_rev};
abrupt step jumps in Sec.~\ref{const_g}; and a control parameter ramp
varying linearly in time in Sec.~\ref{time_dependent_g}. Finally, we
conclude with some remarks connecting our results to those of Buks
and Roukes~\cite{BR} who swept the frequency rather than the driving
strength.

\section{BCL amplitude equation}
\label{sec:bcl}

Lifshitz and Cross~\cite{LC} modeled the array of coupled nonlinear
resonators that was studied by Buks and Roukes~\cite{BR} using the
equations of motion
\begin{align}
\ddot{u}_{n} & + u_{n} + u_{n}^{3} -
\frac{1}{2}\epsilon(\dot{u}_{n+1} - 2\dot{u}_{n} +
\dot{u}_{n-1})\nonumber\label{eom}\\
& +\frac{1}{2}\bigl[\Delta^{2}+\epsilon h\cos(2\omega_{p}t)\bigr](u_{n+1}%
-2u_{n}+u_{n-1})\nonumber\\
& -\frac{1}{2}\delta^{1/2}\bigl[(u_{n+1}-u_{n})^{2}(\dot{u}_{n+1}-\dot{u}%
_{n})\nonumber\\
& -(u_{n}-u_{n-1})^{2}(\dot{u}_{n}-\dot{u}_{n-1})\bigr]=0,
\end{align}
where $u_n(t)$ describes the deviation of the $n^{th}$ resonator from
its equilibrium, with $n=1\ldots N$, and fixed boundary conditions
$u_{0}=u_{N+1}=0$. Detailed arguments for the choice of terms
introduced into the equations of motion are discussed in
Ref.~\cite{LC}.  The terms include an elastic restoring force with
both linear and cubic contributions (whose coefficients are both
scaled to 1), a dc electrostatic nearest-neighbor coupling term with a
small ac component responsible for the parametric excitation (with
coefficients $\Delta^{2}$ and $\epsilon h$ respectively), and linear
as well as cubic nonlinear dissipation terms. The dissipation in the
system is assumed to be weak, which is used to define two small
expansion parameters $\epsilon\ll1$ and $\delta\ll1$ by setting the
linear damping rate to $\epsilon$ and the nonlinear damping
coefficient to $\delta^{1/2}$, with a square root for later
convenience. The driving amplitude is then expressed as $\epsilon h$,
with $h$ of order one, in anticipation of the fact that parametric
oscillations at half the driving frequency require a driving amplitude
which is of the same order as the linear damping rate~\cite{Landau}.
Both dissipation terms are taken to be of a nearest neighbor form,
motivated by the experimental indication that most of the dissipation
comes from the electrostatic interaction between neighboring beams.

In order to treat the system of equations (\ref{eom}) analytically,
BCL introduced a continuous displacement field $u(x,t)$, and slow
spatial and temporal scales, $X=\epsilon x$ and $T=\epsilon t$. They
tried a solution in terms of a pair of counter-propagating plane
waves, oscillating at half the drive frequency,
\begin{eqnarray} \label{uAnsatz}\nonumber
u(x,t)&=& \epsilon^{1/2}
\bigl[\left(A_+(X,T)e^{-iq_px}+A_-^*(X,T)e^{iq_px}\right)e^{i\omega_p
t} \\
&+&  c.c.\bigr] + \epsilon^{3/2}u^{(1)}(x,t,X,T)+\ldots,
\end{eqnarray}
where the asterisk and $c.c.$ denote complex conjugation, and $q_p$
and $\omega_p$ are related through the dispersion relation
\begin{equation}
  \label{eq:dispersion}
  \omega_p^2 = 1 - 2\Delta^2\sin^2\left(\frac{q_p}{2}\right).
\end{equation}
By substituting this ansatz (\ref{uAnsatz}) into the equations of
motion (\ref{eom}) and applying a \emph{solvability condition} on the
terms of order $\epsilon^{3/2}$, BCL obtained a pair of coupled
amplitude equations for the counter-propagating wave amplitudes
$A_\pm$. A linear analysis of these equations shows that at the
critical drive amplitude $h_{c}=2\gamma\omega_p$ a particular linear
combination of the two counter-propagating waves obtains a positive
growth rate, forming a standing wave pattern, while the growth rate of
the orthogonal linear combination remains negative. This implies that
a single amplitude equation should suffice at onset, describing this
standing wave pattern.

At this point it is natural to define a reduced driving amplitude $g$
with respect to the critical drive $h_c$ at onset by letting
$(h-h_{c})/h_{c}\equiv g\delta$, and to introduce a second ansatz,
\begin{eqnarray}\label{Bansatz}
\left(%
\begin{array}{c}
  A_+ \\
  A_- \\
\end{array}%
\right)&=&\delta^{1/4}
\left(%
\begin{array}{c}
  1 \\
  i \\
\end{array}%
\right) \hat{B}(\hat\xi,\hat\tau)+
\delta^{3/4}\left(%
\begin{array}{c}
  w^{(1)}(X,T,\hat\xi,\hat\tau) \\
  v^{(1)}(X,T,\hat\xi,\hat\tau) \\
\end{array}%
\right)\nonumber\\&+&
\delta^{5/4}\left(%
\begin{array}{c}
  w^{(2)}(X,T,\hat\xi,\hat\tau) \\
  v^{(2)}(X,T,\hat\xi,\hat\tau) \\
\end{array}%
\right),
\end{eqnarray}
where $\hat\xi=\delta^{1/2} X$ and $\hat\tau=\delta T$. Substitution
of this ansatz allows one to obtain the correction of the solution at
order~$\delta^{3/4}$
\begin{equation}\label{w1v1Sol}
\begin{split}
&\left(\begin{array}{c}
  w^{(1)} \\
  v^{(1)} \\
\end{array}%
\right)=\frac1{4\omega_p\sin^2(q_p/2)}\\
&\times\left(\Delta^2\sin\left(q_p\right)\frac{\partial \hat{B}}{\partial
    \hat\xi} + 9i|\hat{B}|^2\hat{B}\right) \left(
\begin{array}{c}
  1 \\
  -i \\
\end{array}%
\right),
\end{split}
\end{equation}
after which a solvability condition applied to the terms of order
$\delta^{5/4}$, and a rescaling of all the physical quantities, yield
an equation for the scaled field $B(\xi,\tau)$ of the form
\begin{eqnarray} \label{BampEq}\nonumber
\frac{\partial B}{\partial\tau} &=& gB +
\frac{\partial^{2}B}{\partial\xi^{2}} + i\frac{2}{3}
\left(4|B|^{2}\frac{\partial B}{\partial\xi} +
  B^{2}\frac{\partial B^{*}}{\partial\xi}\right)\\
&- &2|B|^{2}B - |B|^{4}B.
\end{eqnarray}
This is the BCL amplitude equation. It is governed by a single control
parameter, the reduced drive amplitude $g$, and captures the slow
dynamics of the coupled resonators just above the onset of parametric
oscillations. The reader is encouraged to consult Ref.~\cite{BCL} for
a more detailed account of the derivation of the BCL equation, as well
as a detailed list of all the scale factors leading to the final form of
the equation.

\section{Single-mode solutions of the BCL amplitude equation}
\label{sec:single}

The simplest nontrivial solutions of the BCL amplitude equation are
steady-state single-mode extended patterns, given by
\begin{equation}\label{SingleMode}
B(\xi,\tau) = b_k e^{i(\varphi-k\xi)},
\end{equation}
with $b_k$ and $\varphi$ both real. This solution, when substituted
back into (\ref{w1v1Sol}) and (\ref{Bansatz}), and then into
(\ref{uAnsatz}), yields single-mode standing-wave parametric
oscillations at half the drive frequency, whose explicit form is given
in Appendix A. The original boundary conditions $u(0,t)=u(N+1,t)=0$
constrain the phase $\varphi$ to be $\pi/4$ or $5\pi/4$, and constrain
the wave numbers of the spatial pattern to have the quantized values
of $q_m = m\pi/(N+1)$, with $m=1,\ldots, N$.

BCL showed that the first single-mode pattern to emerge as the
zero-state becomes unstable is that whose wave number $q_m$ is closest
to the wave number $q_p$ that is determined by the drive frequency
$\omega_p$ through the dispersion relation~(\ref{eq:dispersion}). This
determines the value of the scaled wave number in the single-mode
solution~(\ref{SingleMode}) to be
\begin{equation}\label{eq:k-zero}
k_0 =  \left(m - q_p \frac{N+1}{\pi}\right) \Delta Q_N,
\end{equation}
where $m$ is the integer closest to $q_p(N+1)/\pi$, and $\Delta Q_N$,
whose explicit value is given in Appendix A, tends to zero as the size
$N$ of the array of resonators tends to infinity.  In this paper we
are interested in secondary transitions as the initial single-mode
state of wave number $k_0$ becomes unstable with respect to the growth
of other single-mode states, whose wave numbers we label as
\begin{equation}
  \label{eq:k-n}
  k_n \equiv k_0 + n\Delta Q_N.
\end{equation}

\begin{figure}
\includegraphics[width=0.9\columnwidth]{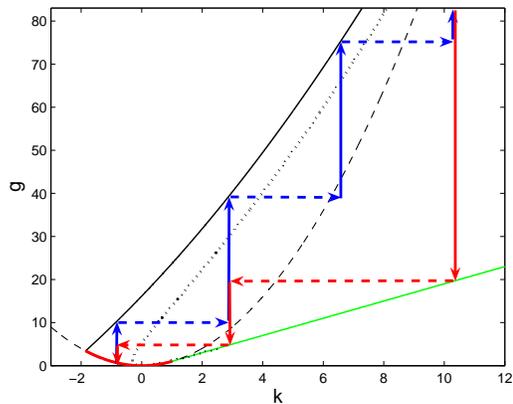}%
\caption{\label{StbBalloon}
  (Color online) Stability boundaries of the single-mode solution
  (\ref{SingleMode}) of the BCL amplitude equation (\ref{BampEq}) in
  the $g$~{\it vs.}~$k$ plane.  Dashed line: neutral stability curve
  $g=k^{2}$. Dotted line: stability boundary of the single-mode
  solution~(\ref{SingleMode}) for a continuous spectrum
  ($Q\rightarrow0$).  Solid lines: stability boundary of the
  single-mode solution for $N=92$ and the parameters $\Delta=0.5$,
  $q_{p}=73\pi/101$, and $\epsilon=\delta = 0.01$ (giving $k_0\simeq
  -0.81$ and $\Delta Q_N \simeq 3.70$).  Black line: the value of $g$
  for which the eigenvalue $\lambda_{g,k}(\Delta Q_N)$ turns positive.
  Red line: the lower bound for $k<1$, $g=k^{2}$.  Green line: the
  lower bound for $k>1$, the locus of saddle-node bifurcations
  $g=2k-1$.  Vertical and horizontal arrows mark the secondary
  instability transitions shown in Fig.~\ref{three_type_sweep} and
  discussed in Sec.~\ref{quasi_rev}.}
\end{figure}

In steady state, the relation between the magnitude $b_k$ and the
wave number $k$ is found by substituting (\ref{SingleMode}) into
(\ref{BampEq}) and setting the time derivative to zero, to give
\begin{equation}\label{Bsteady}
b_k^2 = (k-1) + \sqrt{(k-1)^2 + (g-k^2)}\geq 0,
\end{equation}
along with a negative square-root branch which is always unstable
against small perturbations~\cite{BCL}, as can be verified by the
analysis below.  Linearization of the BCL amplitude equation
(\ref{BampEq}) shows that the zero state with $B(\xi,\tau)=0$---which
is a solution of~(\ref{BampEq}) for any value of $g$---is stable
against the formation of single-mode patterns with wave number $k$ as
long as $g<k^2$. The neutral stability curve $g=k^2$ is plotted as a
dashed parabola in Fig.~\ref{StbBalloon}. Furthermore, for $k<1$ the
bifurcation from the zero state to that of single-mode oscillations is
supercritical, occurring on the neutral stability curve, while for
$k>1$ it is subcritical, with a locus of saddle-node bifurcations
located along the line $g=2k-1$ (shown in Fig.~\ref{StbBalloon} as a
solid green line), where the square root in (\ref{Bsteady}) is exactly
zero.

The stability of a single-mode solution (\ref{SingleMode}) of wave
number $k$ against an Eckhaus transition to a different single-mode
solution of wave number $k\pm Q$ is found by performing a linear
stability analysis of solutions of the form
\begin{equation}\label{BStability}
B(\xi,\tau)=b_ke^{-ik\xi}+\left(\beta_+(\tau)
e^{-i(k+Q)\xi}+\beta_-^*(\tau)e^{-i(k-Q)\xi}\right),
\end{equation}
with $|\beta_\pm|\ll1$. When the larger of the two eigenvalues
describing the growth of such a perturbation, which is given
by~\cite{BCL}
\begin{eqnarray}\label{lambda}\nonumber
    \lambda_{g,k}(Q) & = & 2b_{k}^{2}(k-1-b_{k}^{2})-Q^{2}\\\nonumber
    &+ &\frac{2}{3} \left[3Q^{2}(k - b_{k}^{2})(3k - 5b_{k}^{2})\right.\\
    &&+ \left.9b_{k}^{4}(k - 1 - b_{k}^{2})^{2}\right]^{1/2},
\end{eqnarray}
becomes positive the single-mode solution of wave number $k$ undergoes
an Eckhaus instability with respect to different single-mode solutions
of wave numbers $k\pm Q$\footnote{Note that for the positive square-root
branch (\ref{Bsteady}) $b_{k}^{2}(k-1-b_{k}^{2})<0$, implying that
$\lambda_{g,k}(0)=0$, while for the negative square-root solution
$b_{k}^{2}(k-1-b_{k}^{2})>0$, implying that $\lambda_{g,k}(0)>0$. As a
consequence the positive square-root solution (\ref{Bsteady}) is
stable with respect to small perturbations with the same wave number
$k$, while the negative square-root solution is unstable.}.

For an infinite number of oscillators the Eckhaus instability forms
the upper boundary of the stability balloon of the single-mode
solutions, and also the lower boundary for $k<5/2$. For $k>5/2$ the
lower boundary is the saddle node bifurcation line. For a finite
number of oscillators, restricting $Q$ to be an integer multiple of
$\Delta Q_N$ in (\ref{lambda}) slightly shifts the Eckhaus
instability lines. The upper Eckhaus boundary is shifted to larger
values of $g$. The nature of the lower instability
boundary now depends on the number of resonators in the array through
$\Delta Q_{N}$, as well as on the wave number $k$ \cite{BCL}. For
$k<1$ the lower boundary will be the Eckhaus instability curve if
$|k|>\Delta Q_{N}/2$, and the neutral stability curve otherwise. From
(\ref{eq:k-zero}) and (\ref{eq:k-n}) we find that the only wave
number to satisfy $|k|<\Delta Q_{N}/2$ is $k_{0}$, which means that
upon decreasing $g$ the $k_{0}$ solution undergoes a continuous
transition to the zero state. For $k>1$ the lower boundary will be
the Eckhaus instability curve if $1<k<(5-3(\Delta Q_{N}/2)^{2})/2$,
and the line of saddle node bifurcations otherwise. For $\Delta Q_{N}>2$
(for the parameters used throughout this paper this corresponds to
$N<172$) there is no portion of Eckhaus instability on the lower
boundary, which is the neutral stability curve if $k<1$ and the
saddle node bifurcation curve if $k>1$. These stability boundaries
are shown in Fig.~\ref{StbBalloon} for an infinite system and for a
system of $N=92$ resonators, which is discussed next.

\section{Quasistatic sweeps of the control parameter}
\label{quasi_rev}

\begin{figure}
\includegraphics[width=1\columnwidth]{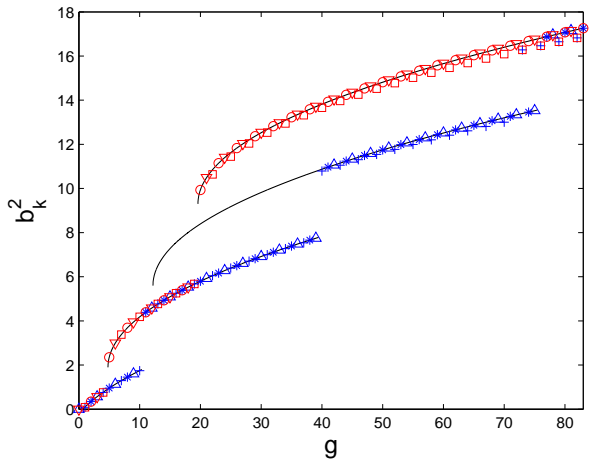}%
\caption{\label{three_type_sweep}
  (Color online) The amplitude of single-mode oscillations as a
  function of the reduced drive amplitude $g$. The parameters are the
  same as in Fig.~\ref{StbBalloon}. Solid lines show the analytical
  values~(\ref{Bsteady}) of the amplitudes $b^{2}_{k}$ of the modes
  $k_{n}$ (for $n=0,\ldots, 3$). Note that the $k_0$ mode bifurcates
  supercritically, whereas all the other modes start at a saddle-node
  bifurcations. All modes terminate at the values of $g$ for which
  they become Eckhaus unstable. Symbols show numerical calculations,
  where blue is used for upward sweeps of $g$ and red is used for
  downward sweeps, as follows: (a) $+$s and $\Box$s show upward and
  downward sweeps, respectively, of the original equations of
  motion~(\ref{eom}) of the coupled resonators. (b) $\bigtriangleup$s
  and $\bigtriangledown$s show upward and downward sweeps,
  respectively, of the BCL amplitude equation~(\ref{BampEq}). (c) $*$s
  and $\circ$s show upward and downward sweeps, respectively, of the
  truncated mode expansion equations~(\ref{a-eqn}) for the seven modes
  $b_{-3}$ to $b_{3}$.  }
\end{figure}

We begin by taking a close look at the switching that occurs between
single-mode patterns~(\ref{SingleMode}) of different wave numbers
$k_{n}$ as the control parameter---the reduced drive amplitude
$g$---is varied quasistatically. We examine a typical situation,
which is depicted within the stability balloon of single-mode
solutions, shown in Fig.~\ref{StbBalloon}. Parameters are chosen such
that the initial pattern happens to have a wave number
$k_0\simeq-0.81$, which corresponds to the array of $N=92$ nonlinear
resonators oscillating at its $m=67$ mode. Because $k_0<1$ we expect
the pattern to grow supercritically from the zero state as the
control parameter is gradually increased from $g=0$.  The sequence of
expected secondary transitions to single-mode patterns of wave
numbers $k_n$ can be understood with the help of the vertical and
horizontal lines drawn within the stability balloon. As $g$ reaches a
value of about 10, the initial $k_0$ pattern undergoes an Eckhaus
instability to a pattern of wave number $k_1\simeq 2.90$.  As this
occurs in the solution~(\ref{SingleMode}) of the amplitude
equation~(\ref{BampEq}) the pattern of the array of nonlinear
resonators~(\ref{eom}) switches from the $67^{th}$ mode to the
$68^{th}$ mode via a single phase slip, in which the number of nodes
in the standing-wave pattern increases exactly by one.  With the
continuing increase of the control parameter $g$ the secondary
pattern eventually undergoes another Eckhaus transition to $k_2\simeq
6.60$ ($m=69$), followed by a further Eckhaus transition to
$k_3\simeq 10.30$ ($m=70$).

Upon decreasing the value of the control parameter $g$ back to zero,
the $k_3$ pattern remains stable down to its saddle-node bifurcation
at a value of $g$ just below 20. As we further decrease $g$, the
$k_{2}$ wave number is skipped and the $k_{1}$ wave number appears,
even though the control parameter is varied quasistatically. This
transition is discussed in detail below. The single-mode pattern of
wave number $k_1$ eventually reaches its saddle-node bifurcation value
and is replaced by the $k_0$ pattern.

This sequence of secondary transitions, which is expected for such a
quasistatic upward sweep of the control parameter followed by a
quasistatic downward sweep, is verified numerically in
Fig.~\ref{three_type_sweep}. Solid curves show the analytical
values~(\ref{Bsteady}) of the amplitudes $b_{k}^{2}$ of the modes
$k_{n}$ ($n=0\ldots 3$), plotted in the region in which the
corresponding single-mode solutions~(\ref{SingleMode}) are stable.
Superimposed symbols show the numerical solution of both the original
equations of motion~(\ref{eom}) for $N=92$ resonators, and the BCL
amplitude equation~(\ref{BampEq}), for a quasistatic sweep of the
control parameter from $g=0$ up to $g=85$ and back down to $g=0$.  We
note that in order to satisfy the boundary conditions when integrating
the BCL amplitude equation, both the $\delta^{1/4}$ and $\delta^{3/4}$
terms of Eq.~(\ref{Bansatz}), when substituted into the expression
$A_{+}e^{-iq_{p}x}+A_{-}^{*}e^{iq_{p}x}$ in Eq.~(\ref{uAnsatz}), must
be set to zero separately at the boundaries.  This yields a pair of
conditions on $B$ and its derivative, at the boundaries, of the form
\begin{eqnarray}
  &&Be^{-iq_{p}x}-iB^{*}e^{iq_{p}x} = 0, \\
  &&\frac{\partial B}{\partial \xi}e^{-iq_{p}x}+i\frac{\partial
  B^{*}}{\partial \xi}e^{iq_{p}x} = 0. 
\end{eqnarray}

To study the actual process of an Eckhaus transition as it takes
place, we expand the general solution of the BCL amplitude equation in
the linear modes of the array
\begin{equation}\label{multimode}
  B(\xi,\tau)=\sum_n b_{n}(\tau)e^{i(\varphi_{n}-k_{n}\xi)},
\end{equation}
where $k_{n}$ is defined in~(\ref{eq:k-n}), as was done, for example,
in a similar situation by Kramer et al.~\cite{ksz}. Substituting a
truncated mode expansion~(\ref{multimode}) containing a finite number
of modes around $k_0$ into the BCL amplitude equation~(\ref{BampEq}),
allows us to replace this partial differential equation with a finite
number of ordinary differential equations for the coupled mode
amplitudes,
\begin{eqnarray}\label{a-eqn}\nonumber
\frac{\partial b_{n}}{\partial \tau}&= &\left(g - k_{n}^{2}\right)
 b_{n}\\\nonumber
 &+ &2\sum_{m,p}\left(k_p - 1 - \frac{m-n}{3} \Delta Q_N \right)
 b_{m}b_{p}b_{m+p-n}^{*}\\
 &- &\sum_{m,l,p,r}b_{m}b_{l}^{*}b_{p}b_{r}b_{m-l+p+r-n}^{*}.
\end{eqnarray}
To satisfy the boundary conditions, as mentioned above for the
single-mode solution~(\ref{SingleMode}), we take each mode amplitude
to be zero at the boundaries by setting all the phases $\varphi_{n}$
in Eq.~(\ref{multimode}) to $\pi/4$, and take the amplitudes $b_{n}$
to be real, keeping in mind that they can be either positive or
negative.  Note that if all mode amplitudes except $b_0$ are set to
zero we obtain a single equation with $n=m=p=l=r=0$, whose
steady-state solution is the same as the single-mode solution of
BCL~(\ref{Bsteady}).

\begin{figure}
\includegraphics[width=0.9\columnwidth]{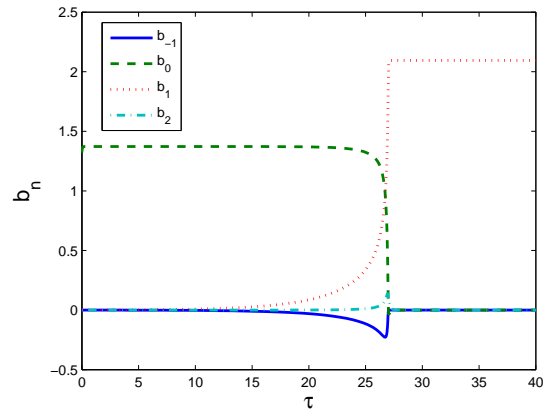}%
\caption{\label{phase_slip} (Color online) Time evolution of the
  amplitudes of the four largest modes that participate in the Eckhaus
  transition from the initial $k_0$ pattern to the $k_1$ pattern,
  obtained by a numerical integration of the seven truncated mode
  equations~(\ref{a-eqn}), for modes $b_{-3}$ to $b_{3}$, using the
  same parameters as in Fig.~\ref{three_type_sweep}. The value of the
  control parameter is changed from $g=10$ to $g=11$ at $\tau=0$,
  causing the initial $k_0$ pattern to become unstable.  The decay of
  the amplitude $b_{0}$ is followed by the rise of $b_{1}$ to its
  expected steady-state value~(\ref{Bsteady}), where it is clearly
  seen that during the transition other modes---including the unstable
  $k_{-1}$ mode---have a non-zero amplitude.}
\end{figure}

\begin{figure}
\includegraphics[width=0.9\columnwidth]{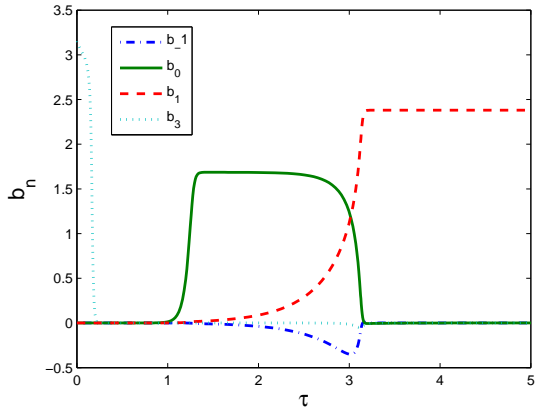}%
\caption{\label{saddle_node} (Color online) Time evolution of the
  amplitudes of the four largest modes as the control parameter is
  changed from $g=20$ to $g=19$, obtained by a numerical integration
  of the seven truncated mode equations~(\ref{a-eqn}), for the
  amplitudes $b_{-3}$ to $b_{3}$, using the same parameters as in
  Fig.~\ref{StbBalloon}.  As the value of $g$ drops below the
  saddle node value of the $k_{3}$ wave number, its amplitude drops
  abruptly to zero. Then, the smallest possible wave number which has
  the largest linear growth rate over the zero solution, $k_{0}$,
  grows to reach its steady-state value~(\ref{Bsteady}). Nevertheless,
  after a short transient the $k_0$ pattern decays through an Eckhaus
  instability and the $k_1$ pattern grows to its steady-state value.}
\end{figure}

As shown in Fig.~\ref{three_type_sweep}, we can capture the sequence
of Eckhaus transitions from the $k_0$ pattern up to the $k_3$ pattern,
and back down to the $k_0$ pattern through the saddle-nodes, by
integrating seven coupled ordinary differential
equations~(\ref{a-eqn}) (calculated symbolically using MATLAB) that
correspond to a truncated mode expansion~(\ref{multimode}) containing
the seven modes from $n=-3$ to $n=3$.  The results agree very well
with those obtained by integrating the full BCL amplitude equation as
well as the original equations of motions for the resonators, yet the
solution in terms of a truncated mode expansion allows us to inspect
the transitions between patterns in greater detail.

We take a closer look at the transient behavior during the first
Eckhaus transition from the initial $k_0$ pattern to the $k_1$ pattern
by plotting the time evolution of the four largest modes during this
Eckhaus transition, as shown in Fig.~\ref{phase_slip}. One can observe
the decay of the unstable mode amplitude $b_{0}$ followed by the
growth of $b_{1}$ to its steady-state value. One can also see that
during the transient the amplitude of the unstable mode $b_{-1}$
becomes non-zero. Its participation in the Eckhaus transition from the
$k_0$ pattern to the $k_1$ pattern is essential, as can be verified by
considering these two modes alone in a truncated expansion.  Limiting
the expansion to $b_0$ and $b_1$ suppresses the Eckhaus transition,
and the $k_0$ pattern remains stable as $g$ exceeds its expected value
for the Eckhaus instability. The Eckhaus transition is observed only
when the $k_{-1}$ mode is included as well, corresponding to the
stability calculation, performed earlier for the state given by
Eq.~(\ref{BStability}).

One might naively expect that the same mechanism causes the transition
from the $k_{3}$ pattern to the $k_{1}$ pattern at $g=19$ through a
double phase slip, however, this is not the case.
Fig.~\ref{saddle_node} reveals the transient processes on a downward
sweep of $g$ just below the saddle node at $g=19$. As $g$ crosses the
saddle node bifurcation point, the amplitude $b_{3}$ drops abruptly to
zero. As can be seen from Eq.~(\ref{a-eqn}), in the zero displacement
state the linear growth rates of the solutions (\ref{multimode}) are
$g-k_{n}^{2}$, so the $k_{0}$ pattern has the largest possible growth
rate and it out-grows the other modes until its amplitude approaches
the steady state value (\ref{Bsteady}).  However, according to the
eigenvalue (\ref{lambda}), at this value of $g$ the $k_{0}$ pattern is
Eckhaus unstable with respect to the $k_{1}$ pattern---notice the
characteristic evolution of the modes around $\tau=3$ in
Fig.~\ref{saddle_node} corresponding to the Eckhaus instability (cf.\ 
around $\tau=25$ in Fig.~\ref{phase_slip}).  Thus the $k_{1}$ mode is
ultimately the selected pattern.

\section{Abrupt change of the control parameter}
\label{const_g}

\begin{figure}
\includegraphics[width=0.9\columnwidth]{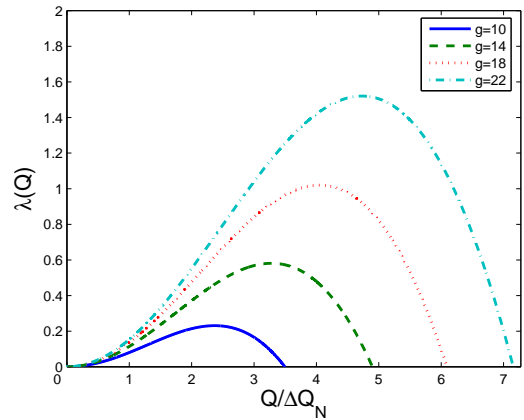}%
\caption{\label{four_lambda} (Color online) The linear growth rate
  $\lambda_{g,k_0}(Q)$ as a function of the wave number shift $Q$,
  plotted for different values of the control parameter $g$.  The
  horizontal axis labels $Q$ in units of $\Delta Q_N$, which for a
  finite system of $N=500$ resonators, with $k_0\simeq-0.075$, has the
  value of $\Delta Q_N\simeq0.69$ (all other parameters are the same
  as in Fig.~\ref{StbBalloon}). The Eckhaus instability of the initial
  $k_0$ pattern occurs for these parameters at $g\simeq5.13$.}
\end{figure}

For a quasistatic increase of $g$ the Eckhaus instability leads to a
single phase slip event and a jump by one of the mode number. This is
no longer always the case for more rapid variations of $g$
\footnote{The behavior when the saddle node instability is encountered
  on decreasing $g$ is expected to be similar for slow or rapid
  variation of $g$, since the dynamics is simply the decay of the mode
  to zero, with the subsequent growth of other modes.}. In this
section we consider the question of pattern selection after an abrupt
jump in $g$, so that single-mode states of different wave numbers
compete with each other after the system is initiated in an Eckhaus
unstable state. We consider a scenario in which the system is
initiated in the $k_0$ single-mode state~(\ref{SingleMode}), after
which the control parameter $g$ is abruptly increased so that the
$k_0$ wave number is no longer stable, while many other wave numbers
become simultaneously stable.  In order to predict the single-mode
pattern that is selected we use our previous expression~(\ref{lambda})
for the eigenvalue $\lambda_{g,k_0}(Q)$ to calculate the linear growth
rate of perturbations of single-mode patterns of wave number $k_0+Q$.
In Fig.~\ref{four_lambda} we plot the $\lambda_{g,k_0}(Q)$ as a
function of $Q$ for four different values of $g$, illustrating the
dependence of the fastest growing wave number $k_{0}+Q_{max}$ on $g$.
The wave number $k_0+Q_{max}$ with the largest linear growth rate
$\lambda_{max}$, which is expected to overcome all other modes, is
obtained by finding the maximum of $\lambda_{g,k_0}(Q)$ as a function
of $Q$, yielding
\begin{eqnarray}\label{Qmax}\nonumber
Q_{max}^{2} & = &(3k_0^2 - 5b_{k_0}^{2}k_0 - 3b_{k_0}^{2} +
  2b_{k_0}^{4})\\
 &\times &\frac{(3k_0^{2} - 11b_{k_0}^{2}k_0 + 3b_{k_0}^{2} +
  8b_{k_0}^{4})}{3(k_0 - 
  b_{k_0}^{2})(3k_0 - 5b_{k_0}^{2})},
\end{eqnarray}
and
\begin{equation}\label{lambda_max}
\lambda_{max}=\frac{(3k_0^{2} - 5b_{k_0}^{2}k_0 - 3b_{k_0}^{2} +
  2b_{k_0}^{4})^{2}}{3(k_0 - b_{k_0}^{2})(3k_0 - 5b_{k_0}^{2})},
\end{equation}
where $b_{k_0}$ is the steady-state amplitude of the unstable $k_0$
mode, as given by Eq.~(\ref{Bsteady}), that depends on the actual
value of $g$.

\begin{figure}
\includegraphics[width=0.9\columnwidth]{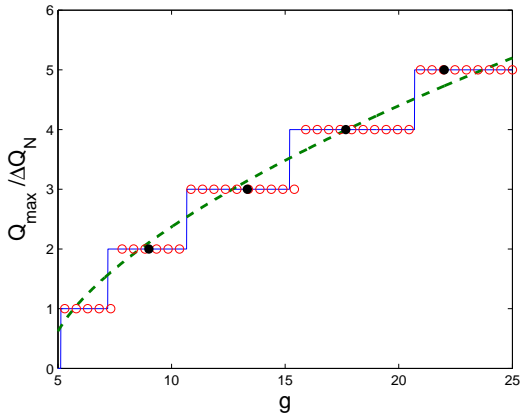}%
\caption{\label{Eckhaus_decay} The wave number shift $Q_{max}$ with the
  maximal growth rate (in units of $\Delta Q_N$) as a function of $g$.
  The blue solid line shows $Q_{max}$ for the same parameters used in
  Fig.~\ref{four_lambda}, and the green dashed line gives $Q_{max}$
  for an infinite system~(\ref{Qmax}). The open red circles are the
  wave number shifts that are observed numerically by integrating the
  BCL amplitude equation~(\ref{BampEq}). The full black circles are
  wave number shifts that are obtained by a numerical integration of
  the original equations of motion~(\ref{eom}), calculated for select
  values of $g$. The numerical calculations are initialized with the
  $k_{0}$ solution and random small-amplitude noise, to initiate the
  growth of competing patterns. We emphasize that the results of the
  numerical solution of the BCL amplitude equation (\ref{BampEq}) are
  not sensitive to the noise amplitude as long as it is sufficiently
  small.}
\end{figure}

For a finite system the selected wave number is expected to be the
$k_{n}$---defined in Eq.~(\ref{eq:k-n})---which has the largest linear
growth rate. The Eckhaus instability is triggered by random
small-amplitude noise. In our finite system the difference between
growth rates of different modes is expected to be sufficiently large
so that by the time nonlinear effects are important, the amplitude of
the fastest growing mode far exceeds those of other destabilizing
Eckhaus modes, and it will reach its steady state value.
Fig.~\ref{Eckhaus_decay} shows $Q_{max}$ for an infinite system and
for a finite system of $N=500$ resonators, where the two curves should
tend to one another as $N\rightarrow\infty$. These predictions for the
selected wave numbers are verified numerically by integrating the BCL
amplitude equation~(\ref{BampEq}), as well as the original equations
of motion~(\ref{eom}) of the coupled resonators. We note that for the
parameters used the stability balloon contains about $10$ modes for
each of the values taken for the control parameter.  For $g=21$, for
example, all modes with wave numbers from $k_{3}$ to $k_{16}$ are
stable.  Finally, by following the amplitude of the growing mode as a
function of time in the numerical solution of the BCL amplitude
equation~(\ref{BampEq}), it is possible to extract the linear growth
rate of the mode numerically. We have done so and found that the
numerically calculated growth rates agree to within $2\%$ with the
analytical values of $\lambda_{g,k_0}(Q_{max})$.

\section{Ramps of the control parameter}
\label{time_dependent_g}

We finish by considering a scenario in which the control parameter $g$
varies smoothly with time---this is often called a control parameter
ramp. To simplify the analysis we consider slow variation in time,
where $dg/d\tau \ll 1$, so that we can use the expressions
(\ref{SingleMode}) and (\ref{Bsteady}), obtained earlier for the
steady-state single-mode solutions of the BCL amplitude equation, with
a simple replacement of the previously constant $g$ by a time
dependent $g(\tau)$,
\begin{eqnarray}\label{t_d_singleMode}
&B(\xi, \tau)=a_n(\tau)e^{i(\varphi-k_{n}\xi)},\\
 \label{t_d_steady} &a_n(\tau)^2 = (k_{n}-1) + \sqrt{(k_{n}-1)^2
+ (g(\tau)-k_{n}^2)}.
\end{eqnarray}
Thus, $a_n(\tau)$ would be the steady-state amplitude of the pattern
with wave number $k_n$ if the drive were varied quasistatically to its
instantaneous value at time $\tau$. For ramps that are not quasistatic
we expect the actual amplitude, which we denote as $\bar{a}_n(\tau)$,
to lag behind its expected value for a quasistatic ramp.  This time
lag phenomenon is known from experiments measuring the heat flow in a
Rayleigh-B$\acute{\textmd{e}}$nard cell \cite{Guenter01}.

\begin{figure}
  \includegraphics[width=0.9\columnwidth]{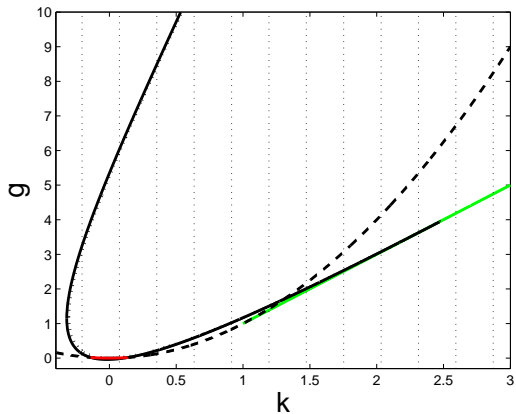}
  \caption{(Color online) Stability balloon for an array of
    $N=1230$ resonators and the same parameters as in
    Fig.~\ref{StbBalloon}, which yields $k_{0}\simeq0.075$ and $\Delta
    Q_{N}\simeq0.28$. The values of $k_{n}$ for $n=-1...10$ are marked
    with vertical dotted lines.  The thick dashed line is the neutral
    stability curve $g=k^{2}$, and the segment in which it is the
    lower boundary $|k|<\Delta Q_{N}/2$ is marked by a solid red line.
    The black dotted and solid curves (which are almost
    indistinguishable) are the Eckhaus boundaries for an infinite and
    a finite system respectively. The solid green line is the saddle
    node $g=2k-1$ which is the lower boundary of oscillations for
    $k>2.47$.}\label{StbBalloon2}
\end{figure}

The specific scenario we examine is one in which the control parameter
increases linearly in time from zero, $g=\alpha\tau$ with
$\alpha\ll1$.  Initially, the system is expected to evolve to the
single-mode state~(\ref{t_d_singleMode}) with wave number $k_0$. As
$g$ increases, this solution becomes Eckhaus unstable and a transition
is expected to a different pattern, which eventually becomes Eckhaus
unstable as well. It is the first of these Eckhaus transitions that we
treat analytically below, as well as test numerically using the BCL
amplitude equation. In order to obtain interesting mode competition,
even for $\alpha\ll 1$, we increase the number of resonators to
$N=1230$, thus increasing the number of stable single-mode solutions
for any particular value of $g$. For such a number of resonators we no
longer perform numerical calculations on the original coupled
equations of motion (\ref{eom}).  The stability balloon for $N=1230$
resonators is shown in Fig.~\ref{StbBalloon2}.  Due to the large
number of resonators, the Eckhaus boundaries for infinite and finite
systems are almost the same.  The dashed vertical lines mark the
values of possible wave numbers $k_{n}$ for $n=-1...10$. For $n=0$
($k\simeq0.075$) the lower boundary of oscillations is the neutral
stability curve, for $n=1...8$ the lower boundary is the Eckhaus
instability curve, and for $n\geq9$ ($k\simeq 2.5$) it is the saddle
node line.

\begin{figure}
  \includegraphics[width=0.9\columnwidth]{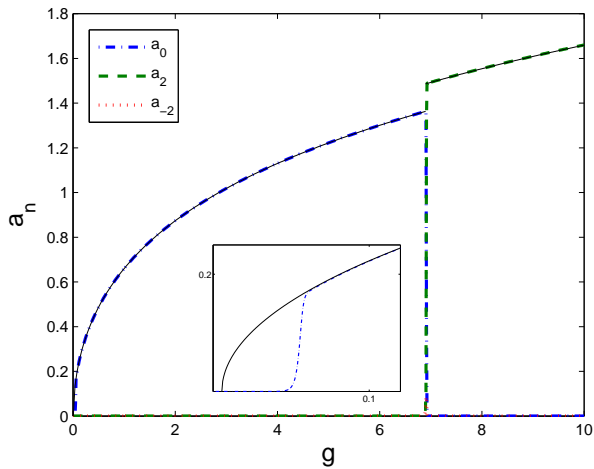}
  \caption{(Color online) The three relevant Fourier amplitudes
    of the numerical solution of the amplitude equation (\ref{BampEq})
    for $g=10^{-4}\tau$ and the same parameters as in
    Fig.~\ref{StbBalloon2}.  The quasistatic values of the
    amplitudes~(\ref{t_d_steady}) are plotted in thin black lines. For
    $\alpha=10^{-4}$ we expect a double phase slip from the $k_{0}$
    pattern to the $k_{2}$ pattern as can be inferred from
    Fig.~\ref{time_dependent}. The inset demonstrates the time lag at
    early times between the actual amplitude of the $k_{0}$ mode and
    its quasistatic value.\label{t_r_phase_slip}}
\end{figure}

\begin{figure}
  \includegraphics[width=0.9\columnwidth]{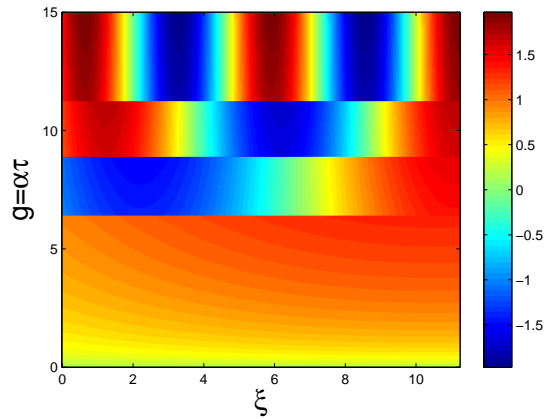}
  \caption{(Color online) The real part of $B(\xi,\tau)$, obtained by
    a numerical integration of the BCL amplitude
    equation~(\ref{BampEq}) for $g=10^{-5}\tau$, using the same
    parameters as in Fig.~\ref{StbBalloon2}.  The initial zero-state
    $B(\xi,0)=0$ evolves into the $k_0$ state, which then undergoes a
    sequence of Eckhaus transitions as $g$ increases in time---the
    first two transitions involve single phase slips, while the third
    involves a double phase slip.}\label{cs_time_ramp}
\end{figure}

A typical response of the system to a linear ramp of the drive
amplitude is shown in Fig.~\ref{t_r_phase_slip} for a ramp rate of
$\alpha=10^{-4}$. One clearly sees the amplitude of the $k_0$ mode,
which forms initially from the zero-state, becoming Eckhaus unstable
around $g\simeq7$ and undergoing a double phase slip to the $k_2$
mode. Thin black lines show the quasistatic values (\ref{t_d_steady})
of the amplitudes of these two modes as a function of $g(\tau)$, while
the blue dot-dashed curve and the green dashed curve show the actual
values of these two amplitudes as obtained by Fourier transforming the
numerical solution of the BCL equation.  The curves are
distinguishable from each other only at very early times, shown in the
inset of Fig.~\ref{t_r_phase_slip}, clearly demonstrating the time it
takes the actual amplitude $\bar{a}_0(\tau)$ to ``catch up'' with the
quasistatic value $a_0(\tau)$, from the zero displacement state. After
this initial time lag the system responds sufficiently quickly so that
the ramp becomes effectively quasistatic.  The only points where the
time dependence of the ramp is still evident are the Eckhaus
instability points at which different ramp rates are expected to lead
to different transitions. A typical sequence of such transitions is
shown in Fig.~\ref{cs_time_ramp} for a ramp rate of $\alpha=10^{-5}$.

\begin{figure}
  \includegraphics[width=0.9\columnwidth]{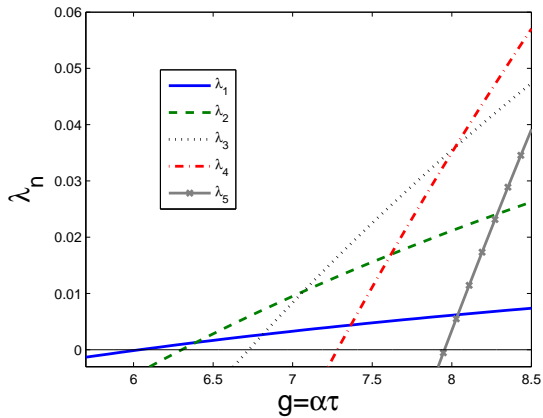}
  \caption{The first five eigenvalues $\lambda_{n}$, plotted as
    a function of $g=\alpha\tau$ using the parameters of
    Fig.~\ref{StbBalloon2}.  As $\lambda_{1}$ turns positive the
    $k_{0}$ solution becomes Eckhaus unstable, but the selected
    pattern depends on the ramp rate $\alpha$ as explained in the
    text.\label{time_ramp_lambda_n_zoom}}
\end{figure}

To analytically predict the first Eckhaus transition, given the ramp
rate $\alpha$, it is useful to introduce a more compact notation for
the eigenvalues (\ref{lambda}), which now depend on time \footnote{For
  slow ramps the instantaneous rate of growth of perturbations is well
  approximated by ignoring the time dependence of the parameter.},
denoting $\lambda_{g(\tau),k_{0}}(n\Delta
Q_{N})\equiv\lambda_{n}[g(\tau)]$. The first five of these eigenvalues
are plotted in Fig.~\ref{time_ramp_lambda_n_zoom} as a function of
$g(\tau)$. Note that as $n$ increases, the corresponding eigenvalue
$\lambda_n$ becomes positive at a later point in time, which we denote
as $\tau_n$, but grows more rapidly than the smaller-$n$ eigenvalues.
At time $\tau_n$ the amplitude of the $k_n$ mode is expected to start
growing from its initial value $\bar{a}_n(\tau_n)$, which in a real
physical system is set by the noise floor. In our analysis below we
take this initial value to be the same as the accuracy of the
numerical routine that is used for integrating the BCL equation. Once
the $k_n$ pattern starts growing it competes with all the other
single-mode patterns with positive growth rates. We expect the
pattern that is eventually selected to be the one whose amplitude is
first to reach the quasistatic value $a_n(\tau)$, given by
Eq.~(\ref{t_d_steady}). Thus, lower-$n$ modes have an advantage for
small ramp rates $\alpha$ because their growth rates become positive
earlier. At higher ramp rates, due to the time-lag phenomenon shown
above, the higher-$n$ modes have an advantage because their
eigenvalues increase more rapidly in time.  This gives rise to an
interesting competition between the possible stable patterns.  A
similar situation was observed in a system described by the stochastic
time-dependent Ginzburg-Landau equation~\cite{TE}.

\begin{figure}
  \includegraphics[width=0.9\columnwidth]{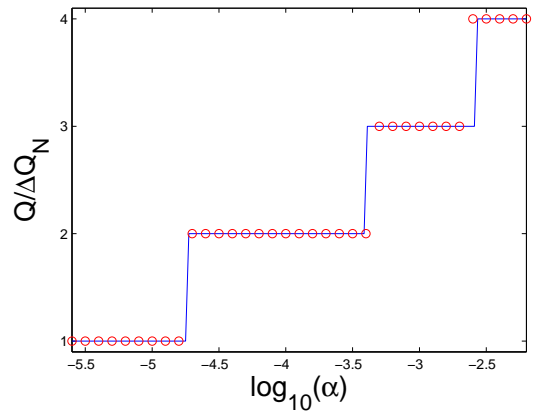}
  \caption{(Color online) The number of phase slips $Q/\Delta Q_{N}$,
    that are observed following the Eckhaus instability of the initial
    $k_0$ pattern, plotted as a function of the ramp rate $\alpha$ for
    a linear ramp of the drive $g=\alpha \tau$. Parameters are the
    same as in Fig.~\ref{StbBalloon2}.  Red circles are the actual
    values observed in the numerical integration of the BCL amplitude
    equation (\ref{BampEq}). The blue line shows the predicted values
    from the linear analysis described in the text, where the initial
    amplitude of each mode, when its eigenvalue becomes positive, is
    taken to be $\bar{a}_{n}(\tau_{n})=10^{-12}$, which is the
    accuracy of the time integration in the numerical solution of
    (\ref{BampEq}).\label{time_dependent}}
\end{figure}

Owing to the slow ramp rates, and the fact that the second eigenvalue
associated with each mode remains negative, we can estimate the growth
of the $n^{th}$ amplitude, in the linear regime, from its initial value
at $\tau_n$ to be
\begin{equation}
\label{time_ramp_amplitude}
\bar{a}_{n}(\tau)=\bar{a}_{n}(\tau_n)e^{\sigma_{n}(\tau,\tau_n)},
\end{equation}
where
\begin{equation}\label{Sigma}
    \sigma_{n}(\tau,\tau_n)=\int_{\tau_n}^{\tau}\lambda_{n}[g(\tau')]d\tau'.
\end{equation}
A comparison of these expressions for $\bar{a}_{n}(\tau)$ for the
different patterns allows us to determine which is the first to reach
its quasistatic value ${a}_{n}(\tau)$, and provides a simple scheme
for predicting the selected pattern following the Eckhaus instability
of the initial $k_0$ pattern. These analytical predictions for the
selected pattern $k_n$ are shown as a function of the ramp rate
$\alpha$ in Fig.~\ref{time_dependent}, and are nicely verified by
numerical integration of the BCL amplitude equation (\ref{BampEq}).
As expected, for small values of $\alpha$ there is a single phase slip
to the pattern with wave number $k_1$.  As $\alpha$ is further
increased this changes to a double phase slip to the $k_2$ pattern
(as demonstrated earlier in Fig.~\ref{t_r_phase_slip}), followed by a
transition to the $k_3$ pattern, and so on.

\section{Conclusions}

We have investigated the sequence of single-mode standing-wave
patterns to be expected in one-dimensional arrays of
parametrically-driven oscillators for time varying drive strengths. An
amplitude equation approach provides a general treatment in terms of a
universal stability diagram on a plot with scaled versions of the
driving strength and wave number as axes. This immediately shows the
type of instability that will be encountered on varying parameters,
and gives qualitative insights on the mode jumps to be expected. For
example, for quasistatic parameter variations, we find that the jump
in the mode number is always unity if the control parameter is
increased so that the Eckhaus instability operates, but larger jumps
are often seen if the control parameter is decreased so that a saddle
node bifurcation occurs. For more rapid increases in the control
parameter larger jumps in the mode number may also occur, and can be
predicted simply from the eigenvalue equation for the Eckhaus
instability. We give explicit results for examples of an abrupt jump
and a slow temporal ramp in the control parameter. In all cases we
checked, simulations of the original oscillator equations of motion
confirm the results based on the amplitude equation.

In the Buks and Roukes experiments on parametrically-driven oscillator
arrays~\cite{BR} which motivated this study the frequency of the drive
was swept, rather than the strength of the driving.  Since the drive
frequency is involved in setting the wave number of the resonant mode
that goes unstable, and these two parameters are involved in a
complicated way in the expressions for the scaled drive and wave
number variables (see BCL), it is more difficult to display the
variation on the scaled stability plot of Fig.~\ref{StbBalloon}.  For
quasistatic or abrupt variations this is immaterial, since the
behavior is determined by where the stability boundary is crossed in
the former case, or the relationship of the final point on the
stability plot to the stability boundaries in the latter case. These
can be estimated quite easily, so that the behavior for quasistatic or
abrupt jumps can be readily predicted. In particular we expect single
mode jumps for quasistatic variations of the drive frequency in the
sense leading to a crossing of the Eckhaus boundary, and the
possibility of larger mode jumps for quasistatic sweeps in the reverse
direction. This is qualitatively consistent with the numerical results
of Lifshitz and Cross~\cite{LC} for a numerical model of 67
parametrically driven resonators, who found more jumps in the solution
on decreasing the frequency than on increasing it. (In the experiments
of Buks and Roukes only upward frequency sweeps were performed.) A
more detailed comparison with experiment would require a better
knowledge of the parameters of the MEMS or NEMS devices so that the
frequency variation could be mapped onto the reduced stability diagram
of the amplitude equation. In future experiments, upward and downward
sweeps of the strength of the driving would provide a more direct
comparison with the theory we have developed.

\section*{Acknowledgments}

This work was funded by the U.S.-Israel Binational Science Foundation
(BSF) through Grant No.~2004339, the U.S. National Science Foundation
under Grant No.~DMR-0314069, and the Israeli Ministry of Science and
Technology.

\appendix

\section{Explicit form of single-mode solutions}
\label{sec:explicit}

When substituting the single mode solution~(\ref{SingleMode}) back
into~(\ref{w1v1Sol}) and~(\ref{Bansatz}), and then
into~(\ref{uAnsatz}), one obtains extended single-mode standing-wave
parametric oscillations at half the drive frequency, whose explicit
form is given by
\begin{eqnarray}\label{second_order}\nonumber
u(x,t) &\simeq &\epsilon^{1/2}\delta^{1/4} 4S_b b_{k}
  \sin(q_{m}x) \bigl[\cos(\pi/4 - \omega_{p}t)\\\nonumber
&+ &\tan(\alpha) \sin(\pi/4 - \omega_{p}t)\bigr]\\\nonumber
&= &\epsilon^{1/2}\delta^{1/4}4S_b b_{k}
  \sin(q_{m}x)(1 + \tan^{2}(\alpha))^{1/2}\\
&\times &\cos(\pi/4-\omega_{p}t-\alpha),
\end{eqnarray}
where we have defined
\begin{equation}
  \label{eq:tanalpha}
  \tan(\alpha) \equiv \delta^{1/2} \frac43 \omega_{p}
  \sin^{4}\left(\frac{q_{p}}{2}\right) \left(b_{k}^2 - k\right),
\end{equation}
and the scale factor
\begin{equation}
  \label{eq:scalefactor}
  S_b \equiv \frac{4}{3\sqrt3} \omega_p
  \sin^3\left(\frac{q_p}{2}\right).
\end{equation}
To satisfy the boundary conditions $u(0,t)=u(N+1,t)=0$, the wave
numbers $q_m$ satisfy the equation
\begin{equation}
\label{eq:q-quantization}
 q_{m}=\frac{m\pi}{N+1}=q_{p}+\frac{k\pi}{\Delta Q_{N}(N+1)},
\end{equation}
where
\begin{equation}
  \label{eq:Qmin}
  \Delta Q_N = \frac{1}{\epsilon\delta^{1/2}}
  \frac{3\Delta^{2}\sin(q_p)}{16\omega_{p}^2
  \sin^{6}\left(\frac{q_p}{2}\right)} \frac{\pi}{N+1}.
\end{equation}

%% Create the reference section using BibTeX:
\bibliography{eyal}
\end{document}